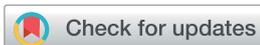

Check for updates

# Density functionals with asymptotic-potential corrections are required for the simulation of spectroscopic properties of materials†

Musen Li, [a] Rika Kobayashi, [b] Roger D. Amos, [b] Michael J. Ford [c] and Jeffrey R. Reimers [*ac]

Five effects of correction of the asymptotic potential error in density functionals are identified that significantly improve calculated properties of molecular excited states involving charge-transfer character. Newly developed materials-science computational methods are used to demonstrate how these effects manifest in materials spectroscopy. Connection is made considering chlorophyll-*a* as a paradigm for molecular spectroscopy, 22 iconic materials as paradigms for 3D materials spectroscopy, and the $V_N^-$ defect in hexagonal boron nitride as an example of the spectroscopy of defects in 2D materials pertaining to nanophotonics. Defects can equally be thought of as being "molecular" and "materials" in nature and hence bridge the relms of molecular and materials spectroscopies. It is concluded that the density functional HSE06, currently considered as the standard for accurate calculations of materials spectroscopy, should be replaced, in most instances, by the computationally similar but asymptotically corrected CAM-B3LYP functional, with some specific functionals for materials-use only providing further improvements.

## Introduction

The asymptotic potential reflects the energy required to take an electron from a system of interest and remove it to the surrounding vacuum. The most commonly used computational method for evaluating spectroscopic properties of molecules and materials, density-functional theory (DFT), can be implemented using functionals embodying many different levels of theory, from the local-density approximation (LDA), to generalised gradient approximations (GGA), to hybrid functionals that mix long-range Hartree–Fock exchange with local exchange, to range-corrected hybrid functionals designed to realistically represent the asymptotic potential, and beyond. This work focuses on the critical effects caused by the difference between basic LDA, GGA, *meta*-GGA, and hybrid functionals and functions with asymptotic correction: their ability to reliably calculate properties of the excited states of molecules and materials. If a spectroscopic transition involves electron transfer from one location to another, then the asymptotic potential error directly manifests in the calculated transition energies.

Two decades ago, the critical role of the asymptotic potential was recognised in regard to the evaluation of molecular spectroscopic properties.[1–23] A significant benefit from this was a gained ability to simulate excitation energies and exciton couplings in photosynthetic reaction centres.[24–26] Even for transitions not involving significant electron transfer, the asymptotic potential error can induce large errors in properties such as calculated reorganization energies and their partitioning into the Huang–Rhys factors (electron-vibration coupling constants) that control spectral bandshapes as well as photochemical and photophysical reaction rates.[27–29] This occurs as such properties are controlled by the details of the entire excited-state manifold, not just the excited state of primary interest.

That such effects are also important for materials was recognised early on,[30] but functionals involving asymptotic correction were not implemented into appropriate software packages. This situation changed recently with the implementation of the CAM-B3LYP asymptotically-corrected functional into both the Car–Parrinello Molecular Dynamics (CPMD) package[31] (gamma-point only) and the Vienna *Ab initio* Simulation Package (VASP)[32] (full implementation). In parallel, new density functionals appropriate only to materials are being developed that link the asymptotic-potential correction to the evaluation of the dielectric constant.[33–35] These methods have greatly improved the modelling of dielectric properties, as well as providing for improved predictions of band gaps. Such approaches lack the general applicability of established methods such as CAM-B3LYP, however, and involve extensive choices made between method options and possibly also their parameterisation. More broadly, many asymptotically corrected functionals have now been developed, including recent efforts,[23,36–44] reflecting many and varied design choices. Whilst we consider mostly CAM-B3LYP herein, the elucidated basic principles are expected to be qualitatively representative of all asymptotically corrected functionals.

[a] *International Centre for Quantum and Molecular Structures and Department of Physics, Shanghai University, Shanghai 200444, China. E-mail: Jeffrey.Reimers@uts.edu.au*

[b] *ANU Supercomputer Facility, Leonard Huxley Bldg. 56, Mills Rd, Canberra, ACT, 2601, Australia*

[c] *University of Technology Sydney, School of Mathematical and Physical Sciences, Ultimo, New South Wales 2007, Australia. E-mail: Mike.Ford@uts.edu.au*

† Electronic supplementary information (ESI) available: Additional tables and figures, plus the geometries of all the structures used plus basic characterization is provided. See DOI: 10.1039/d1sc03738b

This work focuses on the basic understanding as to what asymptotic-potential correction does and why it is essential for studies on both molecules and materials. To do this, we focus on a system that can be regarded as being either a "molecule" or a "material": the defect spectroscopy of hexagon boron nitride (h-BN). This example is taken from the nanophotonics research field,[45–49] an area that embeds into solid-state environments spatially localized properties, with applications ranging from quantum networks and quantum information to spin-photon interfaces.[49–52] Indeed, defects in h-BN have been associated with single-photon emission (SPE)[53–56] and optically detected magnetic resonance (ODMR).[57,58] Historically, only defects displaying magnetic properties have had their chemical natures determined,[57,59–65] but recently came the first characterization based only on observed spectral properties.[66] This was made possible through extensive experimental characterization of composition, combined with computational spectroscopic predictions. In general, the role of computation has been important in all h-BN defect assignments.[61]

For the calculation of the spectroscopic properties of defects in 2D materials, many difficult issues arise with both DFT and *ab initio* wavefunction approaches, with no method that is currently practical able to predict transition energies to within the desired accuracy of say ± 0.2 eV for all possible scenarios.[67] Of tested methods, the most generally reliable approach has been CAM-B3LYP,[7–9] which gives results within 0.5 eV of those from computationally demanding *ab initio* methods.[67]

For both 2D and 3D materials, one of the most widely used density functionals is HSE06,[68,69] a method that also embodies range separation,[3] but uses this feature instead to enhance computational efficiency by removing the Hartree-Fock contributions at long range.[69] Hence a key deficiency of the original PBE functional,[70] upon which HSE06 is based, is made more prominent. Currently, HSE06 is recognised as the standard for the prediction of band gaps of materials.[71] Nevertheless, for the case of defects in 2D materials, HSE06 has been established to make predictions that deviate significantly from those of CAM-B3LYP and *ab initio* approaches[67,72] for problems that do not overtly involve charge transfer.

The example system considered herein is the $V_N^-$ defect in a h-BN 2D layer. This defect consists of a nitrogen-atom vacancy that is negatively charged; it has been considered for a long time[73,74] as a possible contributor to observed h-BN spectroscopic properties, and of late also as a candidate for explaining some observed[58] ODMR, but always not all calculated and observed properties appear to match. Of interest herein, the orbitals associated with the defect are predicted to lie close to the h-BN conduction band,[64,74] with the result that charge-transfer transitions with low energies will occur. This defect therefore allows for the probing of the effect of the asymptotic-potential error on the properties of materials, in a way that directly links to known analogous molecular properties.

First, Results Section (a) reviews the effects of the asymptotic potential on the perceived spectroscopy of an iconic molecular system, chlorophyll-*a*,[9,10] identifying 5 key effects. Section (b) then reviews how these effects manifest concerning predictions of the lowest-energy transition (bandgap) of simple materials,[32] materials historically perceived[71] as being well-treated without the need for asymptotic-potential correction. Comparison of predicted properties from modern materials-only functionals with asymptotic-potential correction to those of generally applicable functionals such as CAM-B3LYP are also reviewed therein, with the results contrasted to those of hybrid functionals including HSE06, as well as some other GGA and *meta*-GGA approaches.

This understanding is then used to interpret calculated properties of the spectroscopic manifold of $V_N^-$. We consider in Sections (c) and (d) molecular models of the defect (Fig. 1), and then 2D periodic models in Section (e) (Fig. 1). Calculations on model compounds for defects in 2D materials show very rapid convergence of calculated electronic properties with respect to increasing sample size for transitions localised within the defect.[65,66,72] Of significance, model-compound calculations have also been shown to converge to the same results as obtained using analogous calculations on 2D periodic defect models.[72,75] Nevertheless, these generic results are not expected to apply to the charge-transfer transitions of $V_N^-$ considered herein. In 3D materials, dielectric screening of long-range interactions can be critical to spectroscopic properties, but this effect does not apply here to the properties of 2D materials.[76] This is a critical factor leading to the found[65,66,72] rapid convergence of both molecular-model and 2D-model spectroscopic calculations with increasing sample size.

## Results

### (a) Overview of how corrections to the long-range potential can become essential for the qualitative interpretation of molecular spectroscopy

Here, ways in which errors in the asymptotic potential manifest to introduce serious qualitative errors in molecular spectroscopy are reviewed, considering the iconic example of chlorophyll-*a*, arguably the world's most important chromophore. The major band systems of this and related molecules (porphyrins, hemes, *etc.*) were characterised by Gouterman in the 1960's,[77,78] in energy order, as: two ~ red transitions, $Q_y$ then $Q_x$, two ~ blue transitions $B_x$ then $B_y$, then a series of weak transitions starting at $N_x$ and $N_y$. In these labels, *x* and *y* indicate absorption polarisation directions in the macrocycle plane. *Ab initio* spectral calculations verify this picture, no matter what approach is taken.[9,79,80] The (weak) $Q_x$ band proved difficult to assign as two *x*-polarised peaks are observed instead of the expected one, leading to two possible assignments that were debated continuously from 1982 (ref. 81 and 82) until 2013.[26] The resolution of this issue is that the Born–Oppenheimer approximation[83] fails for the Q bands, meaning that no classical spectroscopic model (*e.g.*, using Huang–Rhys factors,[84] more general Franck–Condon factors,[85] and/or Herzberg–Teller interactions[86]) can describe the observed spectroscopy: $Q_x$ absorption displays two maxima with a local minimum at the band centre.

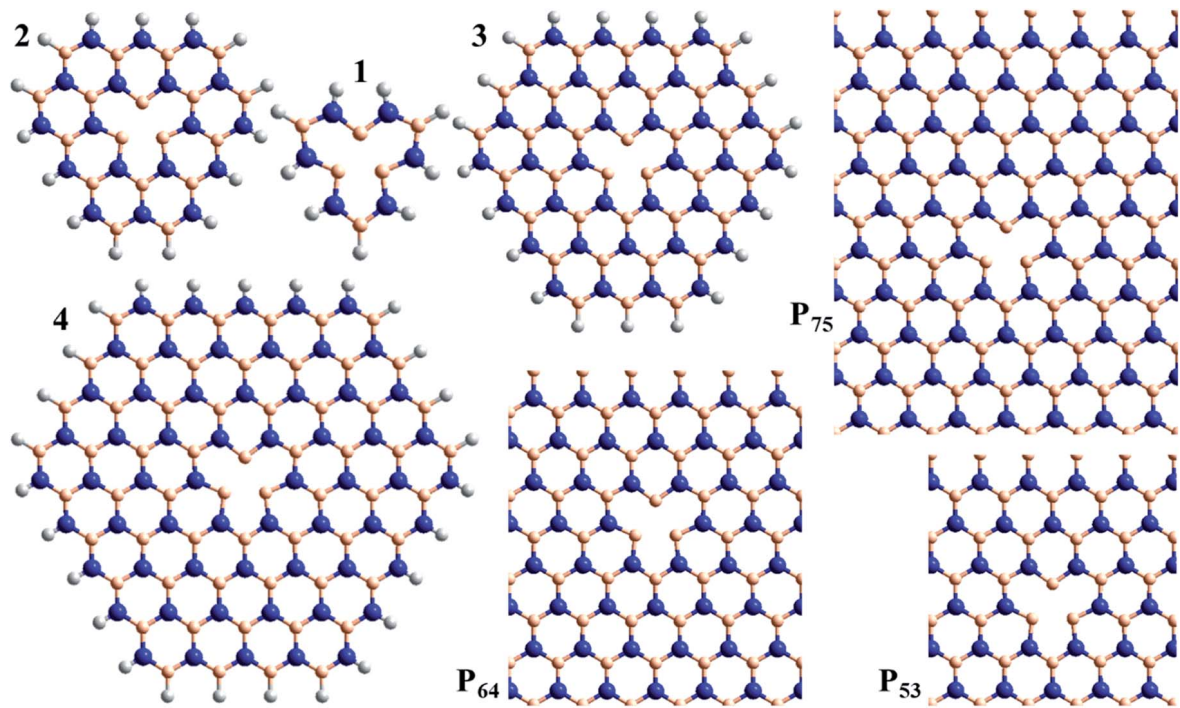

Fig. 1 Atomic models used to consider the spectroscopy of the $V_N^-$ defect in h-BN: compounds **1**, **2**, **3**, and **4**, and 2D-periodic layers $P_{53}$, $P_{64}$, and $P_{75}$ (N-blue, B-beige, H-white).

During the period in which the assignment of the spectrum of chlorophyll-*a* was intensely debated, DFT calculations depicted a scenario in which one of the two *x*-polarised bands was assigned to $Q_x$ and the other to $N_x$, a state unconventionally predicted to occur at low energy.[87,88] The N bands are charge-transfer bands[9] that significantly alter the molecular dipole, and hence are subject to the asymptotic-potential error in DFT. The error was established to be general and result in the underestimation of charge-transfer band energies by several eV.[10] The application of functionals such as CAM-B3LYP with asymptotic-potential correction resulted in predictions of N-band energies consistent with *ab initio* predictions and historical experimental interpretations.[9]

Table 1 gives details of the CAM-B3LYP calculations for chlorophyll-*a*, tracing the effect of long-range correction by comparing these results to analogous B3LYP[89,90] ones and thence to HSE06 results. Highlighted are the differences between calculated and observed values, hampered somewhat as only bounds are available for the N bands and also for one other charge-transfer band involving peripheral-ring ketone excitation. Results for the popular ωB97xD asymptotically corrected density functional[6] are also included and are similar to the CAM-B3LYP results. From the table, it is clear that functions without asymptotic correction like HSE06 and B3LYP give results that are significantly different from those that do.

From Table 1, four key effects of asymptotic-potential correction are extracted and listed in Table 2, along with one additional effect. (1) Asymptotic correction significant increases the bandgap between the HOMO and LUMO orbitals, with (2) a corresponding increase in the exciton binding energy of localised spectroscopic transitions. These two effects largely cancel so that (3) asymptotic correction does not greatly perturb localised transitions. Alternatively, for transitions involving charge transfer (4), the exciton binding energies are enhanced much less and hence charge-transfer transitions increase in energy. This creates order within the excited-state energy manifold. As a consequence (5), the vibronic couplings that occur between excited states become much better described, and hence the excited-state potential-energy surfaces become better described. This means that excited-state equilibrium geometries and spectroscopic reorganisation energies become better described, as well as vibrational line intensities arising from both Franck–Condon (Huang–Rhys) and Herzberg–Teller mechanisms.[27,28,91]

Table 1 Effect of functional variation from HSE06 to B3LYP to CAM-B3LYP (and ωB97xD) on calculated energies[a] (and differences from experiment[b]), in eV, of chlorophyll-*a*

| State | HSE06 | B3LYP | CAM-B3LYP | ωB97xD |
| --- | --- | --- | --- | --- |
| HOMO–LUMO | 2.04 | 2.54 | 4.19 | 5.22 |
| $Q_y$ | 2.21 (0.27) | 2.19 (0.25) | 2.17 (0.23) | 2.16 (0.22) |
| $Q_x$ | 2.40 (0.26) | 2.37 (0.23) | 2.56 (0.42) | 2.58 (0.44) |
| $B_x$ | 3.26 (0.18) | 3.23 (0.15) | 3.49 (0.41) | 3.51 (0.43) |
| $B_y$ | 3.41 (0.03) | 3.38 (0.00) | 3.75 (0.37) | 3.77 (0.39) |
| $N_x$ | 3.16 (<−0.1) | 3.13 (<−0.1) | 3.83 (<0.6) | 3.72 (<0.5) |
| $N_y$ | 3.12 (<−0.1) | 3.10 (<−0.1) | 3.66 (<0.4) | 3.89 (<0.7) |
| CT to carbonyl | 3.25 (<0.0) | 3.21 (<0.0) | 3.83 (<0.6) | 3.83 (<0.6) |

[a] Vertical excitation energies at the CAM-B3LYP/6-31G*/D3(BJ) ground-state geometry.[91] [b] Gas-phase,[92] with the $Q_x$ band centre (an absorption minimum) estimated using established methods:[26] $Q_y$ = 1.94 eV, $Q_x$ = 2.14 eV, $B_x$ = 3.08 eV, $B_y$ = 3.38 eV, $N_x, N_y$ > 3.2 eV.

Table 2 Known effects of asymptotic correction on calculated molecular spectroscopic properties

| Effect | Description |
| --- | --- |
| (1) | Significant increase in the HOMO–LUMO bandgap |
| (2) | Significant increase in exciton binding energies for charge-localised transitions, cancelling the increase in the HOMO–LUMO gap |
| (3) | Effects (1) and (2) combine to produce a significant increase in the relative energy of charge-transfer transitions |
| (4) | Reduction in the differential error in state energies to provide an improved description of the excited-state manifold |
| (5) | Improved description of the shapes of potential energy surfaces, and hence reorganisation energies, Franck–Condon (Huang–Rhys) factors, Herzberg–Teller couplings, *etc.*, as vibronic couplings with the excited-state manifold are much better represented |

### (b) Overview of how corrections to the long-range potential can significantly improve calculated properties of simple materials

Concerning the spectroscopic properties of some simple materials, Table 3 provides a broad summary of the effects of asymptotic-potential correction, comparing known results for up to 22 materials for 6 functionals without asymptotic correction to those for 11 functionals with asymptotic correction. Some new calculations are performed for CAM-B3LYP, as detailed in Methods and ESI.† All results are listed in ESI Tables S1 and S2,† with Table 3 providing a summary in terms of mean-absolute errors (MAD) and maximum errors (MAX).

Firstly, Table 3 shows the MAD error in calculated lattice constants for the methods for which data is available. As previously noted,[32] the error for CAM-B3LYP is much less than that for HSE06, and indeed much less than that for B3LYP, indicating the importance of asymptotic correction. When predicting spectroscopic properties of materials, it may be necessary to optimise structures and hence the apparent reliability of CAM-B3LYP in this regard is of note.

In addition, Table 3 shows MAD and MAX errors for orbital bandgaps. This involves the comparison of calculated bandgaps with those deduced from experiment through estimations of the exciton binding energy and the assumption that the lowest-observed electronic transition corresponds to the HOMO to LUMO transition. The table is divided into methods with and without asymptotic correction: those without it, including HSE06, show large MAD and MAX errors, whilst those with it, including CAM-B3LYP, perform considerably better. This effect is demonstrated most clearly by comparison of the poor results for PBE0 (ref. 33) and B3LYP to the significantly improved results for their asymptotically corrected counterparts, PBE0($\alpha$)[41] and CAM-B3LYP.

There is current interest in the development of *meta*-GGA functionals for the prediction of spectroscopic properties of materials. Available results for the 22 materials considered are summarised in Table 3. The traditional *meta*-GGA functional SCAN[94,105] performs very poorly, with a MAX error of 4.1 eV. In addition, results for the modern mBJLDA functional[94] are shown, with again a poor MAX error of 1.8 eV. This method was selected from amongst a set of 21 new functionals[94] designed in part to reproduce the analysed data, with mBJLDA showing perhaps the best results for this type of data. Similarly, many of the asymptotically corrected functionals listed in Table 3 were also designed considering the analysed data, whereas CAM-B3LYP is an old functional for which no materials properties were used in its construction.

Table 3 MAD and/or MAX errors in lattice parameters, exciton binding energies, and band gaps for up to 22 materials, evaluated using various density functionals (see ESI for details) blocked into those with (bottom) and without (top) asymptotic correction[32–34,41,44,71,93–121]

| Method | Lattice const. MAD/Å | Orbital bandgap MAD/eV | Orbital bandgap MAX/eV |
| --- | --- | --- | --- |
| PBE[34,93] | 0.044 | 2.01[bcf] | 4.9 |
| SCAN[94,105] | 0.021[h] | 1.30[bc] | 4.1 |
| mBJLDA[94] |  | 0.52[g] | 1.8 |
| PBE0 (ref. 33) |  | 0.60[bd] | 2.0 |
| HSE06 (ref. 32, 34 and 93) | 0.033 | 0.86[bf] | 2.7 |
| B3LYP | 0.020[i] | 0.81 | 1.8 |
| CAM-B3LYP[32,a] | 0.004 | 0.33 | 1.1 |
| WOT-SRSH[44] |  | 0.35[e] | 1.2 |
| PBE0($\alpha$)[41] |  | 0.40 | 1.1 |
| SC-hybrid[33] |  | 0.30[d] | 1.5 |
| RSH $\mu$WS[33] |  | 0.20[d] | 1.0 |
| RSH $\mu$TF[33] |  | 0.20[d] | 1.0 |
| RSH $\mu$erfc-fit[33] |  | 0.30 | 1.2 |
| DD-PBEH[34] |  | 0.28[f] | 0.7 |
| RS-DDH[34] |  | 0.23[f] | 0.7 |
| DD0-RSH-CAM[34] |  | 0.30[f] | 1.8 |
| DD-RSH-CAM[34] | 0.031 | 0.25[f] | 1.4 |

[a] This work also, see ESI. [b] Reported[71] for a large data set: PBE 1.2 eV, SCAN 0.9 eV, PBE0 0.6 eV, HSE06 0.5 eV. [c] Excludes data for semiconductors predicted to be metals. [d] Other data set:[33] RSH $\mu$WS 0.29 eV, RSH $\mu$TF 0.30 eV. [e] Other data set after correction for zero-point energy[44] 0.08 eV. [f] Other data set:[34] PBE 1.10 eV, DD-PBEH 0.90 eV, RS-DDH 0.68 eV, DD-RSH-CAM 0.41 eV. [g] Other data set[94] 0.47 eV. [h] Other data sets give 0.030 Å (ref. 94) and 0.025 Å.[105] [i] Other data set, evaluated using numerical functional derivatives, gives[104] 0.053 Å.

Methods like CAM-B3LYP are designed to reproduce directly observable quantities such as oxidation or reduction potentials and spectroscopic transition energies. Such quantities depend on orbital bandgaps, but include Coulomb and exchange interactions involving the associated charges as well. How this works for molecules was described in Table 2, with the results for materials from Table 3 and ESI Table S2† clearly reflecting effect (1): HOMO–LUMO bandgaps are significantly increased in going from B3LYP to CAM-B3LYP or from PBE0 to BPE0($\alpha$).

Effect (2) depicts analogous changes to exciton binding energies, the prediction of which in materials CAM-B3LYP has been reported to be five times more reliable than HSE06,[32] making effect (3) also important. Nevertheless, in materials such as silicon, no charge transfer is involved. Hence effect (2) has only a minor effect that is accurately predicted by CAM-B3LYP,[32] with the consequence that CAM-B3LYP overestimates the orbital bandgap and spectroscopic transition energy by 1.1 eV (Table 3 and ESI Table S2†). Silicon is an unusual and (therefore) important material that has spawned modern generations of asymptotic corrected functionals (Table 3) that focus on the correct description of its dielectric properties.[33,34,41,43,44]

A contrasting example in ESI Table S2† is LiF, for which both effects (2) and (3) are critical. For LiF, CAM-B3LYP calculations predict an orbital bandgap of 13.97 eV and an exciton-binding energy of 1.29 eV, placing the optical transition energy at 12.68 eV, close to the observed[114,122] band centre at 12.6 eV. The current best-estimate of the exciton-bonding energy from experimental data[114] (aided by computational input) is 1.6 ± 0.2 eV, leading to a perceived orbital bandgap of 14.2 ± 0.2 eV. Prior to that,[122] the deduced exciton-binding energy was 1.00 ± 0.06 eV, leading to a perceived orbital bandgap of 13.60 ± 0.06 eV. Hence the CAM-B3LYP calculated exciton-binding energy is also in good agreement with the experimental data. In stark contrast, the exciton binding energy calculated by HSE06 is just 0.06 eV, too small by a factor of over 20. Lithium participates strongly in $\pi$ back bonding and so charge-transfer is expected to be a significant aspect of its properties, making essential its treatment by asymptotically corrected methods.

In a third example of note from ESI Table S2,† the orbital bandgap of $\alpha$-Al$_2$O$_3$ was initially estimated spectroscopically[123] to be at 8.8 eV, with improved analyses[106] leading to a revised value of 9.57 eV that is in agreement with conductivity measurements,[107] and an exciton binding energy of 0.13 eV. These later results are in good agreement with CAM-B3LYP predictions of an orbital bandgap of 9.42 eV and exciton binding energy of 0.20 eV. Alternatively, results for expensive state-of-the-art methods such as Bethe–Salpeter[124] (0.40 eV) and another GW-based approach[125] (0.37 eV) are significantly too large, whilst HSE06 predicts just 0.010 eV.

Summarising these three examples, significant differences between experiment and CAM-B3LYP calculated transition energies or orbital bandgaps can arise either (i) owing to shortcomings in the methodology, demanding use of improved modern long-range-corrected methods,[33,34,41,43,44] or (ii), when charge transfer is significant, from shortcomings in experimental data interpretations. Owing to effect (2), asymptotically corrected methods, exampled here by CAM-B3LYP, are appropriate for the evaluation of exciton-binding energies. For LiF and $\alpha$-Al$_2$O$_3$, CAM-B3LYP results agree to experiment to within *ca.* the experimental uncertainty, as has already been noted[32] for diamond, silicon, NaCl, MgO, 2D h-BN, and 3D h-BN. CAM-B3LYP therefore appears suitable for use in experimental data assignment, whereas asymptotically uncorrected methods, exampled here by HSE06, qualitatively fail to depict key spectroscopic quantities.

### (c) The asymptotic-potential error in 2D defects perceived using TD-DFT and EOM-CCSD calculations on model defect clusters

The previous subsection concerned the properties of the lowest-energy transitions of 3D materials, but often the questions of interest concern the assignment of many observed transitions within a manifold and hence the *relative* ordering of excited states becomes the most critical issue. Here, the relative ordering of transitions in model compounds depicting $V_N^-$ defect-associated transitions in h-BN are considered.

Table 4 provides excited-state energies obtained using time-dependent DFT (TD-DFT)[126] for the model compounds **1–4**. TD-DFT is particularly well-suited to the study of defects as it only requires that its reference state be of mostly closed-shell nature, whereas most defect states are open shell,[67] which indeed is the case for $V_N^-$. These energies are shown diagrammatically in Fig. 2(a and b). Model **1** only supports excitations within the defect core, with larger models adding in additional excitations that involve charge transfer to the h-BN conduction band. TD-CAM-B3LYP predicts that most transitions decrease in energy as the model size increases, with convergence quickly established. In contrast, TD-HSE06 predicts that most states continually decrease in energy as ring-size increases. The exceptions to this are the energies of $(1)^1E'$ and $(1)^3E'$, which quickly converge. The electron densities of the key $a_2''$ and $e'$ orbitals involved in these transitions are shown in Fig. 2(d) and are all localized within the defect core. All other transitions, however, involve excitation to conduction-band orbitals, like the $a_1''$ orbital shown in Fig. 2(d) that becomes occupied in the $(1)^1A_2''$ and $(1)^3A_2''$ states. Overall, these results parallel effects (2)–(4) that depict the impact of asymptotic-potential corrections on molecular spectroscopy, see Table 2.

**Table 4** Calculated vertical excitation energies for the $V_N^-$ defect in h-BN, in eV, from the $(1)^1A_1'$ ground state of the model compounds **1–4** (Fig. 1), obtained using time-dependent methods

| | EOM-CCSD | TD-CAM-B3LYP | | | | TD-HSE06 | | | |
|---|---|---|---|---|---|---|---|---|---|
| State | 1 | 1 | 2 | 3 | 4 | 1 | 2 | 3 | 4 |
| $(1)^1E''$ | 3.65 | 3.36 | 3.11 | 3.02 | 2.97 | 3.26 | 3.10 | 3.02 | 2.99 |
| $(1)^1E'$ | 4.82 | 4.53 | 3.79 | 3.64 | 3.63 | 4.25 | 2.78 | 2.21 | 1.85 |
| $(2)^1E'$ | | | 4.09 | 3.85 | 3.74 | | 3.42 | 2.69 | 2.26 |
| $(1)^1A_2'$ | | | 3.54 | 3.77 | 3.62 | | 2.52 | 2.59 | 1.81 |
| $(2)^1A_1'$ | 5.47 | 5.20 | 4.67 | 3.61 | 3.84 | 4.77 | 3.85 | 2.10 | 2.20 |
| $(1)^3E''$ | 3.09 | 2.77 | 2.33 | 2.21 | 2.18 | 2.66 | 2.27 | 2.16 | 2.13 |
| $(1)^3E'$ | 3.65 | 3.19 | 2.73 | 2.62 | 2.58 | 2.99 | 2.47 | 2.18 | 1.85 |
| $(2)^3E'$ | | 5.28 | 3.75 | 3.61 | 3.59 | 4.87 | 2.78 | 2.39 | 2.22 |
| $(1)^3A_2'$ | | | 3.46 | 3.71 | 3.61 | | 2.43 | 2.54 | 1.80 |
| $(1)^3A_1'$ | 4.86 | 4.43 | 4.32 | 3.60 | 3.82 | 3.97 | 3.51 | 2.08 | 2.19 |

The $(1)^1A_2''$ and $(1)^3A_2''$ states are highlighted as, for them, an occupied orbital is localized on the model boundary and hence could be considered as an artefact induced by using molecular cluster models. Indeed, for **2**, all charge transfer bands are by necessity located on the boundary. Models like **3** and **4** support centrally located charge-transfer states, and, for **4**, 17 charge-transfer transitions are predicted by TD-HSE06 at under 3 eV in energy, many of which involve more central destination orbitals. The charge-transfer effect depicted is therefore a robust prediction of the calculations; realistic predictions of its infinite-sample limit are provided later using 2D periodic-model simulations.

Fig. 2(d) lists the orbital numbers of the illustrated orbitals, with #374 being the HOMO orbital and #375 the LUMO. The CAM-B3LYP ordering is indicated and identifies $(1)^1A_2''$ and $(1)^3A_2''$ as resulting from the HOMO to LUMO transition; analogous HSE06 results are similar, see ESI.† Surprisingly, the low-energy states $(1)^1E'$ and $(1)^3E'$ arise from excitations deep into the unoccupied orbital space to orbitals #406 and #407. These are essentially defect orbitals, but their presence within the h-BN conduction band induces some mixing. As is often found for defects,[67] orbital energy differences provide poor indications of state energy differences, particularly when charge transfer is involved. The Coulomb interactions between the electron and the hole involved in the excitation are critical to the energetics, interactions that scale with the asymptotic potential for charge-transfer transitions.

Table 4 also shows transition energies for **1** evaluated using equation of motion coupled cluster (EOM-CCSD) theory,[127] another time-dependent approach considered to provide useful descriptions of the singlet and triplet manifolds of $V_N^-$. It does not suffer from anomalies concerning its description of long-range electrostatic or charge-transfer effects, and its predictions are similar to those of CAM-B3LYP, but sometimes different to those of HSE06.

### (d) The asymptotic-potential error in 2D defects perceived using DFT and up to CCSD(T) calculations on model defect clusters

As an alternative to TD-DFT, we now consider transition energies evaluated by calculating individual DFT energies, evoking the Kohn–Sham[128] and Gunnarsson–Lundqvist[129] theorems, for both the initial and final states. These approaches can only be applied to only a limited selection of states, and is significantly hampered for $V_N^-$ as DFT and many *ab initio* approaches fail owing to excited-state open-shell character. Qualitatively sensible results are obtained for the $(1)^3E'$ and $(1)^3A_2''$ states, and results for **1–4** are listed in Table 5 and illustrated in Fig. 2(c). Used are the CAM-B3LYP and HSE06 density functionals, as well as the *ab initio* approaches: Hartree–Fock theory (HF),[130] second-order Møller–Plesset (MP2) theory,[131] coupled-cluster singles and doubles (CCSD) theory,[132–134] and this perturbatively corrected for triples excitations, CCSD(T).[135] For the local excitation $(1)^3E'$, the *ab initio* methods appear to converge quickly as the treatment of electron correlation is systematically enhanced, giving results close to those of CAM-B3LYP, with HSE06 deviating slightly further. For the charge-transfer excitation $(1)^3A_2''$, CAM-B3LYP deviates from CCSD by 0.4 eV, a large deviation, but one consistent with other worst-case predictions obtained using CAM-B3LYP.[67] On the other hand, HSE06 underestimates this energy by 0.9 eV. This is another example of the previously identified asymptotic-potential effects (2)–(4) from Table 2.

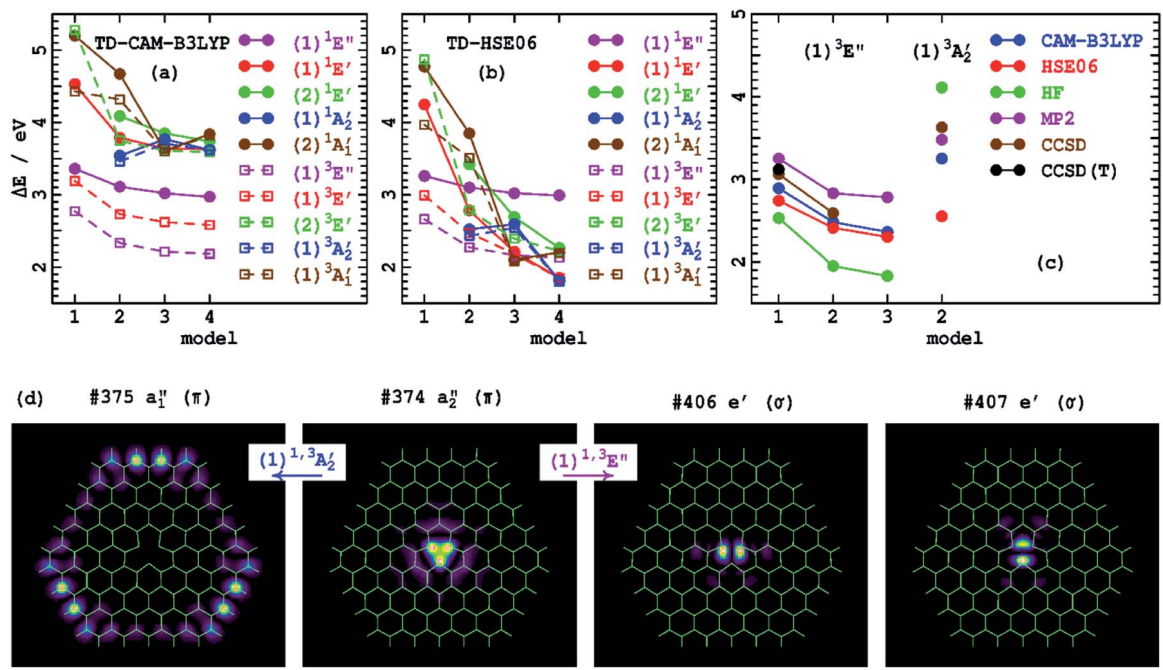

**Fig. 2** Comparison of low-energy singlet and triplet state vertical excitation energies for the $V_N^-$ defect of h-BN (Tables 3 and 4) for varying ring-size models (Fig. 1): (a)-by TD-CAM-B3LYP, (b)-by TD-HSE06, (c)-from DFT calculated state-energy differences. (d) Shows the key orbitals of **4** involved in a localized defect transition and in a charge-transfer transition (CAM-B3LYP orbitals are indicated, analogous HSE06 orbitals are shown in ESI Fig. S1† and are very similar).

Table 5 Calculated vertical excitation energies for the $V_N^-$ defect in h-BN, in eV, from the $(1)^1A_1'$ ground state[a] of the model compounds **1–4** (in $D_{3h}$ symmetry), and the periodic layers $P_{53}$, $P_{64}$, and $P_{75}$ (in just $C_{2v}$ symmetry as necessitated by the boundary conditions, utilizing either $1 \times 1 \times 1$ or $1 \times 2 \times 2$ $k$-points), obtained from state energy differences

|  | $(1)^3E''$ | | | | $(1)^3A_2'$ | $(1)^3B_1$ component of $(1)^3E''$ | | | | | |
| --- | --- | --- | --- | --- | --- | --- | --- | --- | --- | --- | --- |
|  | | | | | | **P53** | **P64** | **P75** | **P53** | **P64** | **P75** |
| Method | 1 | 2 | 3 | 4 | 2 | $1 \times 1 \times 1$ | $1 \times 1 \times 1$ | $1 \times 1 \times 1$ | $1 \times 2 \times 2$ | $1 \times 2 \times 2$ | $1 \times 2 \times 2$ |
| CAM-B3LYP | 2.89 | 2.48 | 2.36 | 2.33[b] | 3.25 | 2.14 | 2.24 | 2.25 | 2.30 | 2.38 | 2.39 |
| HSE06 | 2.74 | 2.41 | 2.30 | 2.27[b] | 2.55 | 1.36 | 1.57 | 1.63 | 1.52 | 1.64 | 1.66 |
| HF | 2.53 | 1.95 | 1.82 | | 4.11 | | | | | | |
| MP2 | 3.25 | 2.83 | 2.72 | | 3.48 | | | | | | |
| CCSD | 3.06 | 2.59 | | | 3.63 | | | | | | |
| CCSD(T) | 3.12 | | | | | | | | | | |

[a] For **1–4**, CAM-B3LYP/6-31G* geometries are used, except for the HSE06 calculations for which HSE06/6-31G* ones are used; for $P_{53}$–$P_{75}$, HSE06 geometries at $1 \times 1 \times 1$ $k$-points are used. [b] Using the cc-pVTZ basis set yields 2.36 eV for CAM-B3LYP and 2.25 eV for HSE06.

### (e) The asymptotic-potential error in 2D defects perceived using DFT calculations on 2D-periodic models

Transition energies can also be obtained for 2D-periodic models from energy differences, but it is difficult to get results for all but the lowest-energy state of each symmetry, and then only reasonable results can be expected if the two states of interest have minimal open-shell character. Results obtained for the $(1)^3B_1$ component of the $(1)^3E''$ state are reported in Table 5, with the electron densities of the two key orbitals involved shown in Fig. 3 for $P_{75}$ (many more frontier orbitals, plus analogous results for $P_{53}$ and $P_{64}$, are shown in ESI†). Owing to the use of boundary conditions in the periodic model, the defect, which should show both $D_{3h}$ and its subgroup $C_{2v}$ symmetry, cannot simultaneously display both 3-fold symmetry elements and the $C_{2v}$ elements, with the utilized lattices retaining only $C_{2v}$ symmetry. In this reduced symmetry, the $(1)^1A_1'$ ground state becomes $(1)^1A_1$, the HOMO orbital changes from $a_2''$ to $b_1$, and we focus on the $a_1$ component of the important $e'$ orbital from Fig. 2(d).

Comparing Fig. 3 and 2(d), we see that HSE06 and CAM-B3LYP both predict that the HOMO orbital of the singlet ground state has very similar form in both the molecular model **4** and the 2D model $P_{75}$, being tightly localized inside the defect. For the 2D model, Fig. 3 shows that this orbital retains its form for spin-up electrons in the lowest triplet state. For the LUMO of the singlet ground state, Fig. 3 also shows similar results from HSE06 and CAM-B3LYP, depicting an orbital that is delocalized over the h-BN, but is most prominent in the box corners so as to maximize distance from the defect. The LUMO orbital becomes occupied in the dominant-spin component of the lowest-energy triplet state, and for it, Fig. 3 reveals very different characteristics predicted by HSE06 and CAM-B3LYP. CAM-B3LYP predicts a localized orbital similar to that predicted by the molecular model **4** (Fig. 2(d)), whereas HSE06 predicts an extremely delocalized orbital. Viewed from the perspective of the ground-state structure, what this indicates is that the electron–hole interaction energies perceived by CAM-B3LYP and by HSE06 are very different. This effect was noted previously for molecular spectroscopy as effect (2) in Table 2, and the HSE06 results are characteristic of significant misrepresentation of the asymptotic potential. As Fig. 3 also indicates that effect (1) holds, with the HSE06 band gap being substantially less than that for CAM-B3LYP, effects (3) and (4) also hold and hence HSE06 significantly underestimates the energies of charge-transfer transitions.

Expanding on the significance of the electron–hole interaction energy, the extended results shown in ESI† indicate that both HSE06 and CAM-B3LYP predict that the $(b_1 \rightarrow a_1)$ excitation has a very large effect on the nature of most of the frontier orbitals. Hence simplistic ideas that the ground-state orbital band structure alone is sufficient to determine excited-state energetics clearly do not apply in this case. Just as the electron–hole interactions generated by the transition dominate the model-compound state properties revealed in Fig. 2, so it also controls the results of the 2D-periodic models. Errors in the asymptotic potential directly manifest in the location-dependence of the electron–hole interaction energy. As a result, HSE06 underestimates the energy of charge-transfer bands. For $V_N^{-1}$, this effect is serious as the lowest-energy transition is incorrectly perceived as being a defect to conduction-band transition instead of an intra-defect transition. So, whereas 2D-periodic and molecular CAM-B3LYP simulations converge to very similar results (Table 5), HSE06 simulations do not.

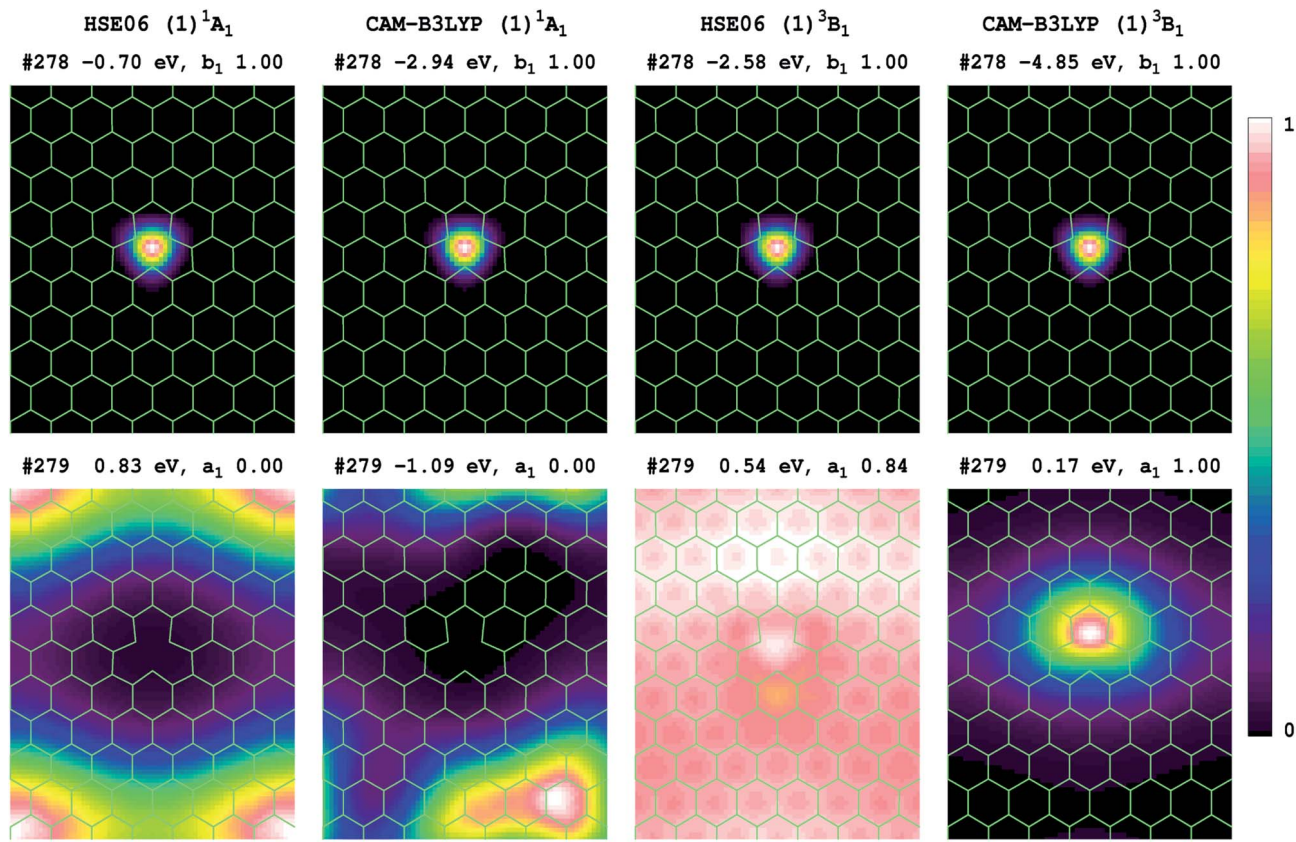

**Fig. 3** Shown for of the $V_N^{-1}$ defect in h-BN are the HOMO and LUMO relative electron densities, along a plane 0.8 Å above the atoms, for orbitals #278 and #279 involved in the $(1)^3B_1$ ($b_1 \rightarrow a_1$) component of the $(1)^3E''$ lowest-energy state. These are based upon 2D model $P_{75}$, using 1 × 2 × 2 $k$-points, and list each orbital energy, symmetry, and occupancy (see SI for many more frontier orbitals and analogous results for $P_{53}$ and $P_{64}$); for the triplet state, only densities for the spin-up component reflecting the electron majority are shown.

## Conclusions

The most important aspects of asymptotic correction to density functionals were identified through reviewing a key application in molecular spectroscopy and listed as five effects in Table 2. Considering just bandgaps of simple materials, effects (1) and (2) were then shown to be significant, despite these materials *not* having dominant charge-transfer character: asymptotic correction increases orbital bandgaps and exciton couplings to provide improved descriptions of the lowest-energy spectroscopic excitation. Effects (3) and (4), concerning the relative ordering of transitions of different charge-transfer types within the excited-state manifold, were then demonstrated to have profound effects on spectral assignment by considering the spectrum of the $V_N^-$ defect in h-BN. When charge transfer is critical, methods without asymptotic correction fail to deliver qualitatively useful results; when charge-transfer character is not critical, asymptotic correction still leads to significant quantitative improvements in calculated properties.

Overall, correcting the asymptotic potential results in significant increases to band gaps, differences that are compensated by analogous increases in exciton binding energies for charge-localised transitions, but left standing for charge-transfer bands. The net effects include: accurate prediction of band gaps, accurate predictions of exciton binding energies, accurate predictions of spectroscopic transition energies, and improved prediction of geometrical properties such as lattice constants. Related known effects include significant improvements to calculated reorganisation energies, Huang–Rhys factors, and chemical and photochemical reaction rates, paving the way for the computational assignment of uninterpreted spectra.

HSE06 should be replaced as the default standard for the accurate and reliable prediction of materials spectroscopic properties by asymptotically corrected functionals. Approaches like CAM-B3LYP are now available in materials-science computational packages such as CPMD[31] and VASP[32] and require no more computational resources than does HSE06.[32] They offer significant improvement for property prediction, in most cases. The exception to this concerns the prediction of the band gap of silicon, an unusual but extremely important material. Density functionals with asymptotic correction, designed for materials only applications, are now also becoming widely available,[33–35] and offer solutions for silicon and related problems, whilst more generally applicable asymptotically correct methods are currently also being devised.[23,36–44] Optimal methods that are independent of dimensionality will provide the opportunity for unified understanding of molecular and materials spectroscopies.

## Methods

As shown in Fig. 1, calculations are performed for the model compounds **1**, **2**, **3**, and **4** of $V_N^-$ that comprise rings of B and N atoms surrounding the nitrogen vacancy, as well as for the model 2D periodic layers $P_{53}$, $P_{64}$, and $P_{75}$. All molecular-model geometries are optimized using CAM-B3LYP/6-31G*, while the layers are optimized using HSE06; full details are provided in ESI.† Only vertical excitation energies are considered. The molecular calculations are performed using Gaussian-16 (ref. 136) B.01 with mostly the 6-31G* basis;[137] convergence of defect spectroscopic properties with respect to basis-set expansion is known to be very rapid.[72] The periodic-cell calculations are performed using VASP 5.4.4 (ref. 32, 138 and 139) (using "PREC = HIGH", "PREFOCK = NORMAL", PBE-PAW pseudopotentials[140] for B3LYP and HSE06 calculations and GW-PAW ones for CAM-B3LYP, with the implied basis sets (NGX, NGY, and NGZ, *etc.*) used for each calculation listed in the ESI†).[32] They are performed using either $1 \times 1 \times 1$, $1 \times 2 \times 2$, or $1 \times 3 \times 3$ representations of the Brillion zone of $(5 \times 3\sqrt{3})R30°$ ($P_{53}$), $(6 \times 4\sqrt{3})R30°$ ($P_{64}$), and $(7 \times 5\sqrt{3})R30°$ ($P_{75}$) unit cells, with lattice vectors depicting an intrinsic h-BN BN bond length of 1.452 Å. Fermi-level smearing is also applied, at an electronic temperature of 0.02 eV. Dipole corrections to mitigate the fictitious interactions between parallel h-BN planes are not used as they are only of order 0.01–0.03 eV. All singlet-state calculations are performed spin restricted, whereas all triplet-state calculations are spin unrestricted. Symmetry analyses[141] of the VASP excited states are listed in the ESI,† along with wavefunction projections that establish the relationship between the ground and excited states.

The new CAM-B3LYP results reported for lattice parameters (ESI Table S1†), orbital bandgaps (ESI Table S2†) and exciton binding energies (text) were also determined using VASP 5.4.4,[32,138,139] with optimised coordinates and calculation details for each material listed in the ESI.† PBE-PAW pseudopotentials[140] were again used for B3LYP and HSE06 calculations, with GW-PAW ones used for CAM-B3LYP.[32] The plane-wave basis sets were truncated at an energy of 850 eV, with many *k*-points sampled until a minimum was obtained in the orbital bandgap, see ESI Table S3.† Exciton-binding energies were determined for vertical excitation by direct evaluation of the energy of the triplet excited states corresponding to the observed singlet transitions, followed by application of (very minor) corrections for the singlet-triplet differences, with details given in ESI Table S4.† The B3LYP geometry optimisations reported in Table 3 were performed using PREC = HIGH and $12 \times 12 \times 12$ *k*-points. These calculations differ from previously reported ones through the use of more modern pseudopotentials and the use of analytical functional derivatives.[32]

## Data availability

Significant data is provided in ESI,† with other information availabe from the authors.

## Author contributions

M. L., assisted by J. R. R., performed the calculations; J. R. R., assisted by M. L. and all others wrote the manuscript; R. K., R. D. A., M. J. F., and J. R. R. designed and supervised the work.

## Conflicts of interest

There are no conflicts to declare.

## Acknowledgements

This work was supported by resources provided by the National Computational Infrastructure (NCI) Australia, as well as Chinese NSF Grant #11674212. Computational facilities were also provided by the ICQMS Shanghai University High Performance Computer Facility. Funding is also acknowledged from Shanghai High-End Foreign Expert grants to R. K. and M. J. F.

## Notes and references